\documentclass[preprint]{aastex}

\makeatletter

\newcommand{\Rmnum}[1]{\expandafter\@slowromancap\romannumeral #1@}
\makeatother

\shorttitle{Modeling the Heliospheric Magnetic Fields}
\shortauthors{Jiang et al.}

\begin{document}

\title{Modeling the Sun's open magnetic flux and the
heliospheric current sheet}

\author{J. Jiang, R. Cameron, D. Schmitt and M. Sch\"{u}ssler}
\affil{Max-Planck-Institut f\"{u}r Sonnensystemforschung, 37191
Katlenburg-Lindau, Germany} \email{jiang@mps.mpg.de}

\begin{abstract}
By coupling a solar surface flux transport model with an
extrapolation of the heliospheric field, we simulate the evolution
of the Sun's open magnetic flux and the heliospheric current sheet
(HCS) based on observational data of sunspot groups since 1976. The
results are consistent with measurements of the interplanetary
magnetic field near Earth and with the tilt angle of the HCS as
derived from extrapolation of the observed solar surface field. This
opens the possibility for an improved reconstruction of the Sun's
open flux and the HCS into the past on the basis of empirical
sunspot data.
\end{abstract}


\section{Introduction}

The Sun's open magnetic flux is the part of its flux which is not
contained in closed loops, but extends into the heliosphere. It is
the source of the heliospheric magnetic field (HMF) whose variations
are an important source of geomagnetic activity \citep{pulk07} and
control the production of cosmogenic isotopes by galactic cosmic
rays \citep{beer00}. A crucial feature of the HMF is the
heliospheric current sheet (HCS), the interface separating the
opposite polarities of the HMF. The tilt angle of the HCS (defined
as the mean of the maximum northern and southern extensions of the
HCS) is a key parameter for the modulation of the flux of galactic
cosmic rays in the inner heliosphere
\citep{kot83,fer04,alan07,heb09}.

At a given distance from the Sun, the HMF has an almost uniform
magnitude in latitude and longitude \citep{bal95}. Therefore, its
radial component near Earth, which has been measured by spacecraft
since the 1960s, faithfully represents the Sun's total open flux
\citep{owens08,lock09}. The tilt angle of the HCS could, in
principle, be measured by multiple spacecraft orbiting at different
heliolatitudes. However, with the exception of the {\em Ulysses}
probe, all measurements of the HMF have been obtained near the
ecliptic plane, so that direct measurements of the tilt angle of the
HCS are not available most of the time. Therefore, such data are
derived by extrapolation of solar surface field maps, such as those
taken at the Wilcox Solar Observatory (WSO) since 1976. This yields
the current sheet distribution at the source surface, where the
field is assumed to become radial \citep[e.g.,][]{hoe82}. In the
inner heliosphere, we may ignore dynamical effects such as the
acceleration of slow plasma and the deceleration of fast plasma that
occur when neighboring parcels of plasma interact \citep{ril02}, so
the field lines are assumed to stay purely radial beyond the source
surface. Under these conditions, the morphology of the HCS may be
inferred from the position of the current sheet at the source
surface.

A limitation of this semi-empirical determination of the HCS tilt
angle arises from the decreasing reliability of the surface field
measurements at higher latitudes and from magnetographic saturation
effects, which cannot be corrected without further assumptions. Flux
transport models based upon observed large-scale magnetic flux
emergence (e.g., in sunspot groups) provide a complementary
possibility to obtain information about the high-latitude surface
fields, which control the open flux and the HCS tilt angle during
solar minimum periods \citep[e.g.][]{wang89a,wang00,mac02}.
\citet{sch06} showed that such an approach reproduces well the HMF
over multiple solar cycles, provided that the heliospheric current
sheet is explicitly included in the field extrapolation
\citep[current sheet source surface method, cf.][]{zha95a}

The present paper serves a two-fold purpose. Firstly, it extends the
study of \citet{sch06} until the current solar minimum period and
includes an explicit account of the HCS and its tilt angle, which
can be compared with the corresponding observational quantities.
Secondly, such comparison provides the validation for the
application of these methods to reconstruct the open flux and HCS
tilt angle backward in time.

The paper is organized as follows. In Sect. 2, we give the
description of the models used. The results are presented in Sect.
3: photospheric flux distributions in Sect. 3.1, the solar open flux
in Sect. 3.2, and the heliospheric current sheet in Sect. 3.3. We
give our conclusions in Sect. 4.

\section{Methods}
\subsection{Surface flux transport model}

The surface flux transport (SFT) model describes the evolution of
the magnetic flux distribution at the solar surface as a combined
result of the emergence of bipolar magnetic regions (BMRs), flux
cancellation, and transport by surface flows. The evolution of the
radially orientated surface field \citep{wan92,sol93,pet09} is
controlled by latitudinal differential rotation and meridional flow,
together with turbulent diffusion due to granulation and
supergranulation \citep[e.g.,][]{wang89b,mac02}.

The SFT uses the radial component of the induction equation in the
form
\begin{eqnarray}
\frac{\partial B_r}{\partial t}=&-&\omega(\theta)\frac{\partial
B_r}{\partial \phi}-\frac{1}{R_\odot\sin\theta}
\frac{\partial}{\partial\theta}\left[\upsilon(\theta)B_r\sin\theta\right]\nonumber \\
&+&\frac{\eta_h}{R_\odot^{2}}\left[\frac{1}{\sin\theta}\frac{\partial}{\partial\theta}
\left(\sin\theta\frac{\partial B_r}{\partial\theta}\right)+
\frac{1}{\sin^{2}\theta}\frac{\partial^{2}B_r}{\partial\phi^2}\right]\nonumber\\
&-&D_r(\eta_r B)+S(\theta,\phi,t),
\end{eqnarray}
where $S(\theta,\phi,t)$ is the source term describing the emergence
of new magnetic flux and $D_r(\eta_r B)$ is the decay term
parameterizing the radial diffusion of the magnetic field
\citep{bau06}. Following \citet{jia09}, the horizontal diffusivity
$\eta_h$ and radial diffusivity $\eta_r$ are taken as
$600~\rm{km^{2}~s^{-1}}$ and $100~\rm{km^{2}~s^{-1}}$, respectively.
For the meridional flow $\upsilon(\theta)$, we adopt the profile
\begin{equation}
\upsilon(\theta)=\left\{
  \begin{array}{l l}
     -\upsilon_0\sin(2.4*(90^{\circ}-\theta)) & 15^{\circ} \leq \theta \leq 165^{\circ} \\
     0 & \textrm{otherwise},
  \end{array}
  \right.
\end{equation}
where $\upsilon_0$ = 11 m s$^{-1}$. This profile is largely
consistent with helioseismic results \citep{giz04}. For the
latitudinal differential rotation $\omega(\theta)$, we use the
empirical profile determined by \citet{sno83}.

The basis for magnetic flux input to the SFT model by emerging BMRs
is the USAF/NOAA sunspot group record
\footnote{http://solarscience.msfc.nasa.gov/greenwch.shtml}. Since a
sunspot group typically appears more than once in the record, we
consider a group only at the day of its maximum area.  Figure 1
gives the monthly number and the latitude distribution (butterfly
diagram) of the emerging sunspot groups for the time interval 1976
-- 2009, which provide the flux input for the SFT model.

The observed sunspot areas have been converted to the areas of the
corresponding BMRs following the procedure of \citet{bau04}. We include
the magnetic flux contained in faculae and plages by employing the
statistical relationship between sunspot area, $A_s$, and facular area,
$A_f$, determined by \citet{cha97},
\begin{equation}
A_f=414+21A_s-0.0036A_s^{2},
\end{equation}
(in units of millionths of the solar hemisphere) and take
$A_{\rm{BMR}}=A_s+A_f$ as area of the corresponding BMR. Since there
is no information about the magnetic polarity in the USAF/NOAA data,
we use Hale's polarity law to infer the polarities of the leading
and following parts of the BMRs. We resolve the ambiguity arising
from the overlap of cycles around activity minima by assuming that
BMRs emerging below $\pm15^\circ$ latitude during the overlap period
belong to the old cycle while all others belong to the new cycle.
The angular separation, $\Delta \beta$ (in degrees), between the
leading and following polarity patches of a BMR is assumed to be
proportional to the square root of the BMR area (given in square
degrees): $\Delta \beta=0.6\sqrt{A_{\rm{BMR}}}$. The angular
separation is separated into latitudinal and longitudinal
components, depending on the BMR tilt angle, $\gamma$, with respect
to the E--W direction. We assume the relation $\gamma=0.15 \lambda$
, which is consistent with observational
results \citep[][Dasi Espuig et al., in preparation]{howa91,siva99}%
\footnote{\citet{sch06} attempted to fix the latitude-dependence of the
          tilt angle independently by calibrating with the
          longitude-averaged surface field integrated over latitude,
          viz.  $\int\mid\int B_r
          d\phi\mid\cos(\lambda)d\lambda/{4\pi}$.  However, some
          exploratory experiments we performed have shown that this
          quantity is rather sensitive to the observed scatter
          of the BMR tilt angles about the latitude-dependent
          mean. Since the effect of this scatter is not significant for
          the quantities that we study in this paper (open flux and
          tilt of the heliospheric current sheet), we do not
          consider BMR tilt angle fluctuations here.}.
Finally, we calibrate the conversion factor between BMR area and
magnetic flux by matching the observed and simulated values of the
disk-averaged unsigned flux density, $B_s = \int\int\mid B_r \mid
d\phi \cos(\lambda) d\lambda /{4\pi}$.

\subsection{Field extrapolation model}

In order to determine the coronal and heliospheric magnetic field
from its source in the photospheric field, a field extrapolation
method is required. For the field distribution on a global scale,
the most widely used approach is the potential field source surface
(PFSS) model \citep{sch69,alt69}. However, the PFSS model (which
includes only volume currents beyond the spherical source surface,
where the field is assumed to become purely radial) does neither
reproduce the latitude-independent radial field found with {\em
Ulysses} \citep{bal95} nor does it match the measured interplanetary
radial field near Earth \citep{sch06}.  The current sheet source
surface (CSSS) model \citep{zha95a,zha95b, zha02}, which explicitly
takes into account the existence of the HCS, does not suffer from
these deficiencies and provides a reasonable match to the measured
quantities \citep{sch06}.

We briefly describe the main features of the CSSS model as follows.
To include the effects of volume and sheet currents, the exterior of
the Sun is divided into three parts, which are separated by two
spherical surfaces, the {\em cusp surface} at $r=R_{\rm{cs}}$, and
the {\em source surface} at $r=R_{\rm{ss}}$ ($\mathbf{R_{\rm{cs}} <
R_{\rm{ss}}}$). In the region inside the cusp surface, the field is
potential. In the region between $R_{\rm{cs}}$ and $R_{\rm{ss}}$,
all flux loops are reconfigured with volume currents and  current
sheets, so that the field becomes completely open. In the region
beyond $R_{\rm{ss}}$, the field is purely radial. Apart from
$R_{\rm{cs}}$ and $R_{\rm{ss}}$, the third adjustable parameter of
the CSSS model is the height scale, $a$, of the horizontal current.
Following \citet{zha02}, we take $a=0.2R_\odot$ in our calculations.
We choose $R_{\rm{ss}}=10.0R_\odot$ according to the estimate of
\citet{mar84}. The value of $R_{\rm{cs}}=1.8R_\odot$ is determined
by the comparison between the magnetic flux at the cusp surface and
the measured radial field from OMNI2 data in Sect. 3.2.

Technically, the opening of the flux beyond $R_{\rm{cs}}$ and the
introduction of the current sheet(s) is carried out by first
calculating a global potential-field extrapolation for the entire
volume above the solar surface. The unsigned radial component of the
field at the cusp surface is then used as the boundary condition for
calculating a magnetic field distribution between $R_{\rm{cs}}$ and
$R_{\rm{ss}}$ of the form in \citet{bog86} and used by
\citet{zha95a}. The orientation of the field lines is then changed
where necessary so that the sign and magnitude of the radial field
at $R_{\rm{cs}}$ is continuous. The distribution of the field
outside $R_{\rm{cs}}$ can then be determined.

\section{Results}

\subsection{Photospheric magnetic field distributions}
The initial condition for the photospheric flux distribution is the
same as the one assumed by \citet{bau04}. It satisfies an
approximate balance between the effects of poleward meridional flow
and equatorial diffusion \citep{van98}. The memory of the system
regarding the initial field depends on the value of the radial
diffusion parameter, $\eta_r$ \citep{bau06}. We use
$\eta_r=100~\rm{km^{2}~s^{-1}}$, which leads to a memory of about 20
yrs. We start all our simulations from the beginning of the
USAF/NOAA sunspot group record in 1874, but consider the results
only for the time period 1976--2009. Therefore, there is no
remaining influence of the initial fields on the results presented
below.

The upper panel of Fig.~2 shows a comparison of observed and
simulated time evolution of the averaged unsigned flux density at
the solar surface, $B_s$ (defined in Sec.~2.1), which has been used
to calibrate the relation between area and magnetic flux of the BMRs
providing the input for the flux transport model. Note that, without
any other parameter adjustment, the model reproduces well the ratio
between the maximum and minimum values as well as the very low
surface flux during the current minimum.

The lower panel of Fig.~2 gives the corresponding time evolution of
the high-latitude surface field, averaged over the caps poleward of
$\pm 75^{\circ}$ latitude. The field amplitude and reversal times
before 2002 are consistent with the observational results given by
\citet{arg02}. For the current minimum, the reported values of the
polar field do not give a consistent picture, reflecting the
difficulties and limitations of polar field measurement.
\citet{pet09} obtained polar fields of 5 -- 6 G by analyzing
photospheric and chromospheric vector polarimetric data obtained
with SOLIS at NSO, which is consistent with our model. On the other
hand, MWO magnetograph \citep{sval05} data and more indirect indices
\citep{schat05} suggest that the present polar field could possibly
be a factor 2 smaller than that during the previous activity minimum
around 1997 (see further discussion in Sect. 3.4).

\subsection{Open flux and near-Earth radial field}
We can calculate the total open flux $\Phi_{\rm{open}}$ resulting from
our extrapolation model by integrating the unsigned radial magnetic
field over the source surface, viz.
\begin{equation}
\Phi_{\rm{open}}(t)=R^{2}_{\rm{ss}}\int\int|B_r(R_{ss},\lambda,\phi)|d\Omega.
\end{equation}
Actually, in the CSSS model the open flux is already fixed at the
cusp surface, $R_{\rm{cs}}$, smaller values of which leading to a
larger amount of open flux. In order to compare with data provided
by near-Earth measurements, we also calculate the longitudinally
averaged unsigned radial field near Earth (at ${r_{\rm E}}=1$~AU and
$\lambda=0$, thus ignoring the slight variation due to the angle of
about 7 degrees between the equatorial plane of the Sun and the
ecliptic plane),
\begin{equation}
    B_{\rm E} (t) = \frac {1}{2\pi}
    \left(\frac {R_{\mathrm ss}}{r_{\mathrm E}}\right)^2
    \int_0^{2\pi} |B_r\left(R_{\mathrm ss},0,\phi,t \right)| \, d \phi \, .
\label{eq:BEq}
\end{equation}
Since extrapolations with the CSSS model reproduce the
latitude-independence of the radial field at $r={r_{\rm E}}$
\citep{sch06}, we have $B_{\rm E} \simeq \Phi_{\rm{open}}/{(4\pi r_{\rm
E}^{2})}$ to a high degree of accuracy, so that $B_{\rm E}$ also
represents the open flux. Similarly, it has been shown that, possibly
after some correction for kinematical effects due to the solar wind, the
measured radial field near Earth can be taken as a reliable proxy for
the total open flux \citep{owens08,lock09}.

Figure~3 shows 27-day averages of $B_{\rm E}$ from our combined
SFT/CSSS model (with $R_{\rm{cs}}=1.8R_\odot$,
$R_{\rm{ss}}=10R_\odot$, $a=0.2R_\odot$, red curve) in comparison
with the measured radial field from OMNI2 data (blue curve)
\footnote{http://omniweb.gsfc.nasa.gov/}. \citet{lock09} have
suggested that the observed data better represent the open flux of
the Sun if a correction of kinematical effects due to the
longitudinal structure of the solar wind is applied. Data modified
in this way are also shown in Fig.~3 (green curve).

The phase relation between the solar activity cycle and the
near-Earth field (and thus the open flux) is well represented by our
model, with $B_{\rm E}$ reaching its peak values $\sim$ 2 -- 3 yr
after activity maximum \citep[cf.][]{mac02,wan02, sch06}. Including
the diffusion in radial direction in the SFT model and the realistic
tilt angles of BMRs with respect to the E--W direction are the main
reasons that lead to the correct phase relation. The only periods
showing a significant disagreement between our model and the data
are the ascending phases of the activity cycles, where the model
values are too low. Possibly the model misses open flux from small
coronal holes at intermediate latitude during this phase. In
principle, this could be corrected by putting the cusp surface
nearer to the Sun during this cycle phase or by assuming a
non-spherical cusp surface. The amplitude of the variation of
$B_{\rm E}$ is also affected by the value of radial diffusion in the
SFT model. However, such parameter tuning would not provide further
physical insight while the overall agreement between model and data
is already encouraging and sufficient for the purposes of this
paper.

\subsection{Heliospheric current sheet (HCS)}
Figure 4 shows the distributions of the magnetic field on the solar
surface, on the cusp surface, and on the source surface during
typical solar minimum and solar maximum conditions in our coupled
SFT/CSSS model. Around solar minimum (left panels), the field at
(and outward of) the source surface assumes a `split-monopole'
structure \citep{bana98}, the HCS separating the two polarities
being located near the equatorial plane \citep{hu08}. Near activity
maximum (right panels), the HCS shows strong excursions in latitude
and additional localized current sheets may occur. Apart from the
current sheets, the field is always largely latitude-independent.
The tilt angle of the HCS is defined by convention as the arithmetic
mean of the maximum northward and southward excursions of the HCS
(see http://wso.stanford.edu/Tilts.html). At the moments shown in
Fig.~4, the HCS tilt angle is $6.5^\circ$ in the activity minimum
period and $71.3^\circ$ in the maximum period.

A comparison between the tilt angle of the HCS resulting from our
SFT model with the values derived by PFSS extrapolations of WSO maps
of the surface magnetic field is shown in Fig.~5a. The results from
our standard CSSS extrapolation (solid red line) do not
significantly deviate from those obtained by a PFSS extrapolation of
the same SFT results with $R_{\rm{ss}}=3.25R_\odot$ (dashed red
line), suggesting that the calculated tilt angle does not
sensitively depend on the extrapolation method (in contrast to the
distribution of open flux). The other two curves in Fig.~5a
represent PFSS extrapolations based on observed photospheric
magnetic fields using two different boundary conditions for the
photospheric magnetic field (see
http://wso.stanford.edu/Tilts.html). Both the `classic' (green
curve) and `new' (blue curve) extrapolations provided by the WSO
website assume that the field is potential between the photosphere
and the source surface. The two models differ in the way the
photospheric field observations are used as the inner boundary
condition and in the height of the source surface. In the `classic'
case the surface observations are taken to correspond to the
line-of-sight component of the potential field, and the source
surface is assumed to be at 2.5 $R_\odot$. In the `new' model, the
observed field is matched to the radial component of the potential
field projected onto the line of sight: the assumption is that the
field in the photosphere is purely radial and becomes potential only
above the surface. Extrapolation for the `new' case exist on the WSO
website for source surface heights of 2.5 $R_\odot$ and 3.25
$R_\odot$. We show the result for 3.25 $R_\odot$ which is claimed to
better match Ulysses data.

The result from our model is consistent with both extrapolations.
Figure~5b shows the difference between the HCS tilt angle derived
from the SFT model with CSSS extrapolation and the other three
cases. The agreement with the WSO `new' method is somewhat worse
than that of the `classic' method, especially around the activity
minimum periods. For physical reasons, it is expected that the `new'
method assuming a purely radial photospheric field should be a
better representation of the real solar situation
\citep{wan92,pet09}. Note, however, that inferring the radial field
strength from the measured line-of-sight component requires quite
large corrections factors at high solar latitudes together with the
larger uncertainties of the high-latitude data. This suggests that
the tilt angles derived from observed synoptic maps during solar
minimum periods should be considered with some caution.

\subsection{The minimum of cycle 23}

The current solar minimum appears to be rather extended, with
particularly low activity levels. There are indications that the
polar field strength is significantly lower than during the two
previous minima \citep[e.g.][]{sval05,schat05,sch08}. A low polar
field strength would be consistent with the small measured values of
the near-Earth interplanetary magnetic field (and thus also the open
flux, see Fig.~3) and the relatively large tilt angle of the HCS
inferred by the `classic' PFSS extrapolation (blue curve in
Fig.~5a). However, the uncertainties in the inferred values of the
polar field are large and the HCS tilt angle determined with the
(presumably more relevant) `new' PFSS model using a radial-field
photospheric boundary condition does not show unusually large
minimum values.

Our SFT model yields a polar field during the present minimum which
is is not significantly weaker than that of the previous minimum
(see Fig.~2). The open flux determined from the model is consistent
with the low OMNI2 measurements during the current minimum; however,
the model results tend to be too low during minima, so that the
present agreement might well be fortuitous. The tilt angle of the
HCS derived from the SFT model decreases to low values during the
current minimum, which is consistent with the `normal' polar field
strength that the model yields. The tilt angles inferred from PFSS
extrapolations based on observed photospheric field distributions
provide a confusing picture (cf. Fig.~5a): while the `classic'
line-of-sight boundary condition leads rather large tilt angles,
which would be in accordance with a weak polar field, the `new'
radial-field boundary condition predicts values not much different
from those during previous cycles. As the `new' method is considered
to be physically more realistic \citep{pet09}, this would dilute the
case for an unusually weak polar field during the present minimum.

In the SFT model, the magnitude reached by the polar field in the
second half of a cycle crucially depends on the tilt angle (with
respect to the East-West direction) of the emerging bipolar magnetic
regions during the cycle. We have assumed the same relationship,
$\gamma=0.15\lambda$, between tilt angle and emergence latitude for
all cycles considered. However, the analysis of sunspot group data
indicates that the factor of proportionality in this relation may
actually vary from cycle to cycle (Dasi Espuig et al., in
preparation). Systematically smaller tilt angles during cycle 23
could lead to a unusually weak polar field during the present
minimum. Alternatively, \citet{sch08} proposed an increased strength
of the diverging meridional flow near the equator as an explanation
for the weak polar field.

\section{Conclusions}
We have simulated the temporal evolution of the Sun's total open
magnetic flux and the heliospheric current sheet (HCS) since 1976 by
coupling a surface flux transport (SFT) model and the current sheet
source surface (CSSS) extrapolation method, using the observed
sunspot groups to provide the magnetic flux input for the model. We
draw the following conclusion from our results:

1) The simulated open flux matches the OMNI2 data quite well,
except for systematically lower values in the ascending phase of the
activity cycle. This is consistent with the view that the solar open
flux is largely determined by the instantaneous photospheric sources.

2) At the source surface, the magnetic field satisfies the `split
monopole' configuration suggested by the {\em Ulysses} out-of-ecliptic
measurements.

3) The temporal variation of the tilt angle of the HCS from the
SFT/CSSS model matches the values derived from potential field
source surface(PFSS) extrapolations of the observed photospheric
magnetic field. The best agreement is found for the `classic'
line-of-sight condition.

4) The conditions during the present minimum period of cycle 23 as
provided by the SFT/CSSS model are similar to those at previous
minima: the polar field has about the same strength as that of cycle
22, the tilt angle of the HCS is small, and the open flux is roughly
at the level of the last minimum. Since 2007, however, the HCS tilt
angle has deviated significantly from the WSO PFSS values with
line-of-sight boundary condition (`classic' case) and approached
those with the `new' radial boundary condition. These results may
well be affected by a systematic variations of the sunspot group
tilts with respect to the E--W direction from cycle to cycle, as
indicated by recent analysis of sunspot observations.

5) In spite of some deviations in detail, the overall agreement of
the model results with observationally inferred values of open flux
and current sheet geometry is encouraging. It opens the possibility
to extend the model backward in time by using the sunspot group
record since 1874. This will be the topic of a subsequent paper.

\acknowledgments Acknowledgments: Y.-M. Wang kindly provided the
observational datasets of the averaged unsigned photospheric field
shown in Fig.~2. M. Lockwood kindly provided the kinematically
corrected OMNI2 data shown in Fig.~3.

%\clearpage

\begin{figure}
\epsscale{.80}\plotone{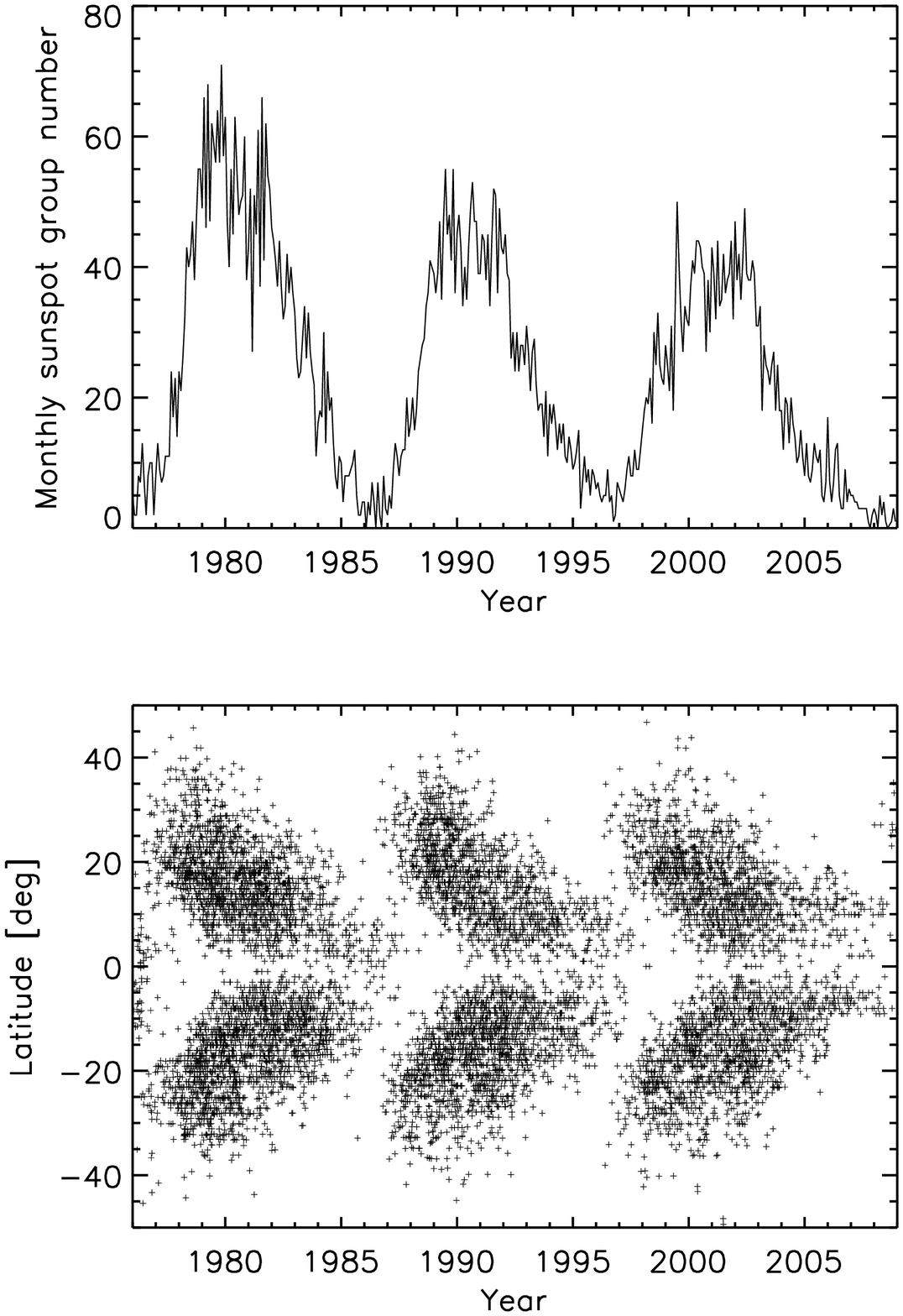} \caption{Number of sunspot groups per
month (upper panel) and time-latitude plot of the emerging sunspot
groups (lower panel) that constitute the input sequence of BMRs to
the SFT model . Data source: USAF/NOAA.}
\end{figure}

%% Here we use \plottwo to present two versions of the same figure,
%% one in black and white for print the other in RGB color
%% for online presentation. Note that the caption indicates
%% that a color version of the figure will be available online.
%%

\begin{figure}
\epsscale{.80} \plotone{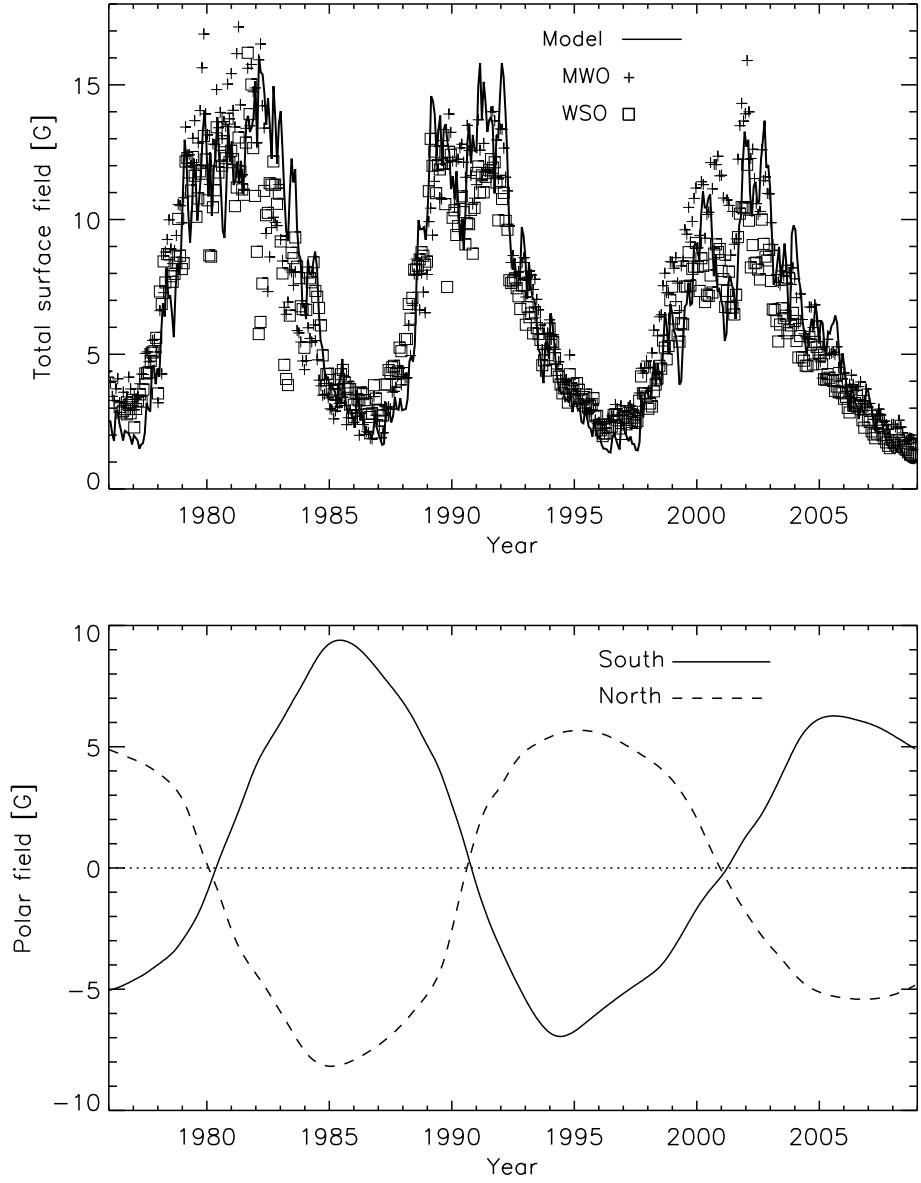} \caption{Upper panel: Averaged
unsigned photospheric field as a function of time. The symbols show
the observed data from the Wilcox and Mount Wilson observatories
(averaged over Carrington rotations). The solid curve represents
27-day averages from the SFT simulation. Lower panel: The temporal
evolution of south (solid curve) and north (dashed curve) polar
fields (averages poleward of $\pm75^{\circ}$ latitude).}
\end{figure}

%% This figure uses \includegraphics to scale and rotate the still frame
%% for an mpeg animation.

\begin{figure}
\plotone{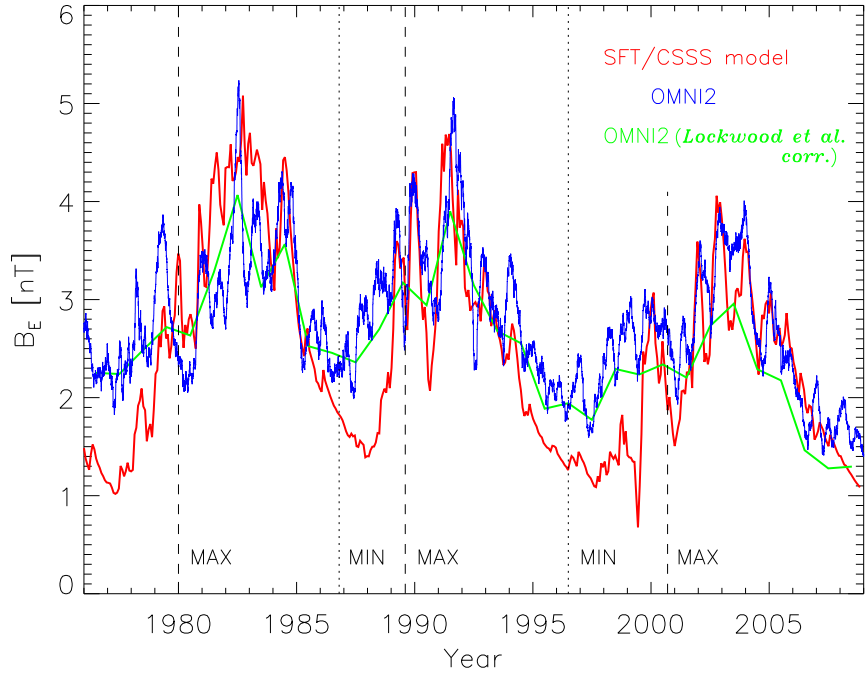} \caption{Temporal evolution of the unsigned radial
field at 1 AU. Shown are the results derived from the SFT/CSSS model
(red curve) in comparison with near-Earth measurements (blue curve).
The latter have been obtained by first averaging the (signed) OMNI2
data over 1-day intervals to remove small-scale fluctuations
\citep{loc06} and then carrying out a 3-month running average of the
unsigned values. The green curve represents yearly averages of the
measurements after applying a kinematic correction \citep{lock09} to
remove effects due to the longitudinal structure in the solar wind.
Dashed and dotted vertical lines indicate the epochs of solar cycle
maxima and minima, respectively.}
\end{figure}

\begin{figure}
\epsscale{0.8} \plotone{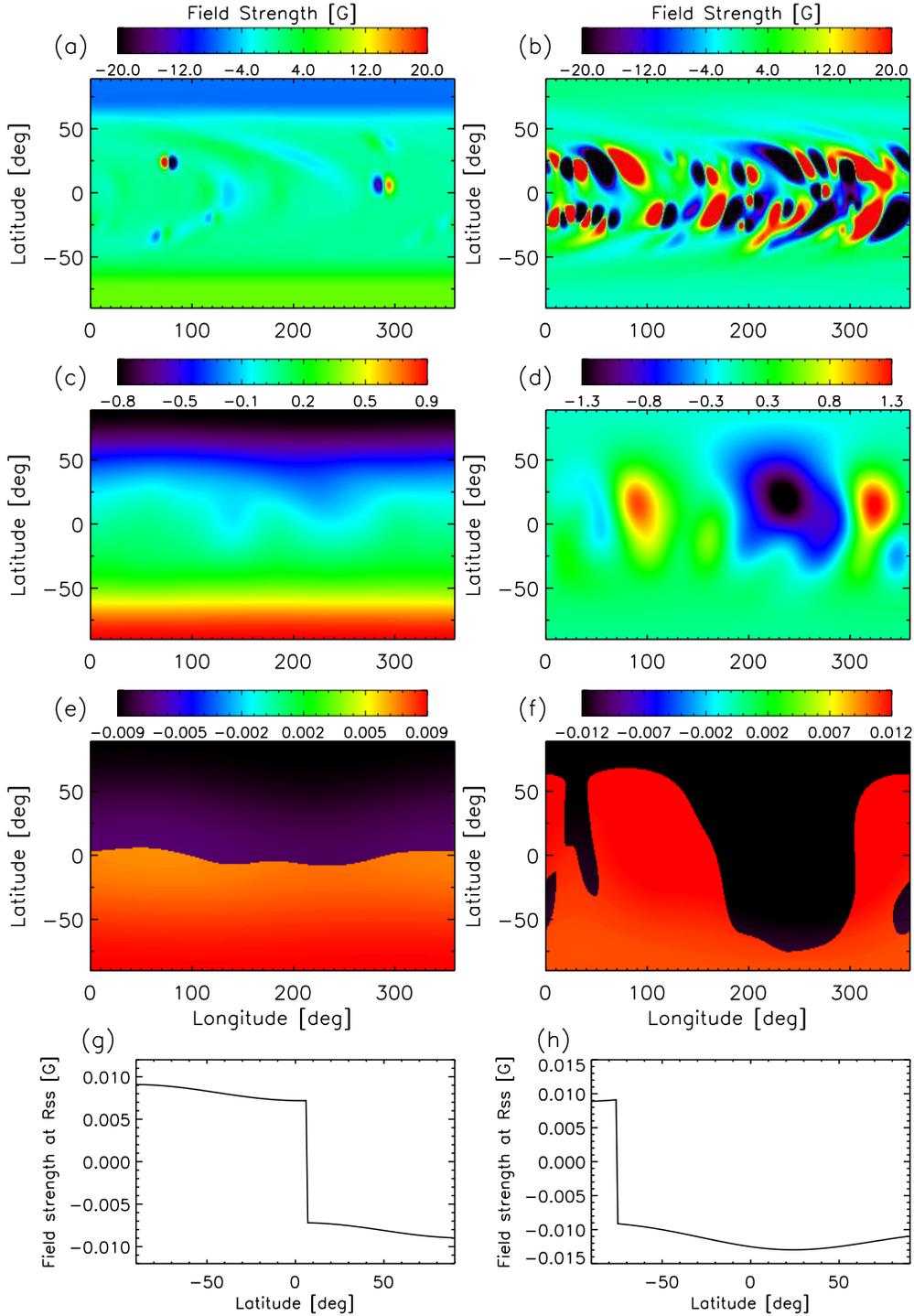} \caption{Magnetic field distributions
at solar minimum (left panels, in 1987.2) and solar maximum (right
panels, in 2000.5) resulting from the coupled SFT/CSSS model. Maps
of the radial field are shown at the solar surface (a,b), at the
cusp surface at $R_{\rm{cs}}=1.8R_\odot$ (c,d), and at the source
surface at $R_{\rm{ss}}=10R_\odot$ (e,f). Panels g and h show
latitude profiles of the field strength at the source surface, taken
at the longitude with the largest latitude excursion of the
heliospheric current sheet (HCS).}
\end{figure}

\begin{figure}
\epsscale{0.8} \plotone{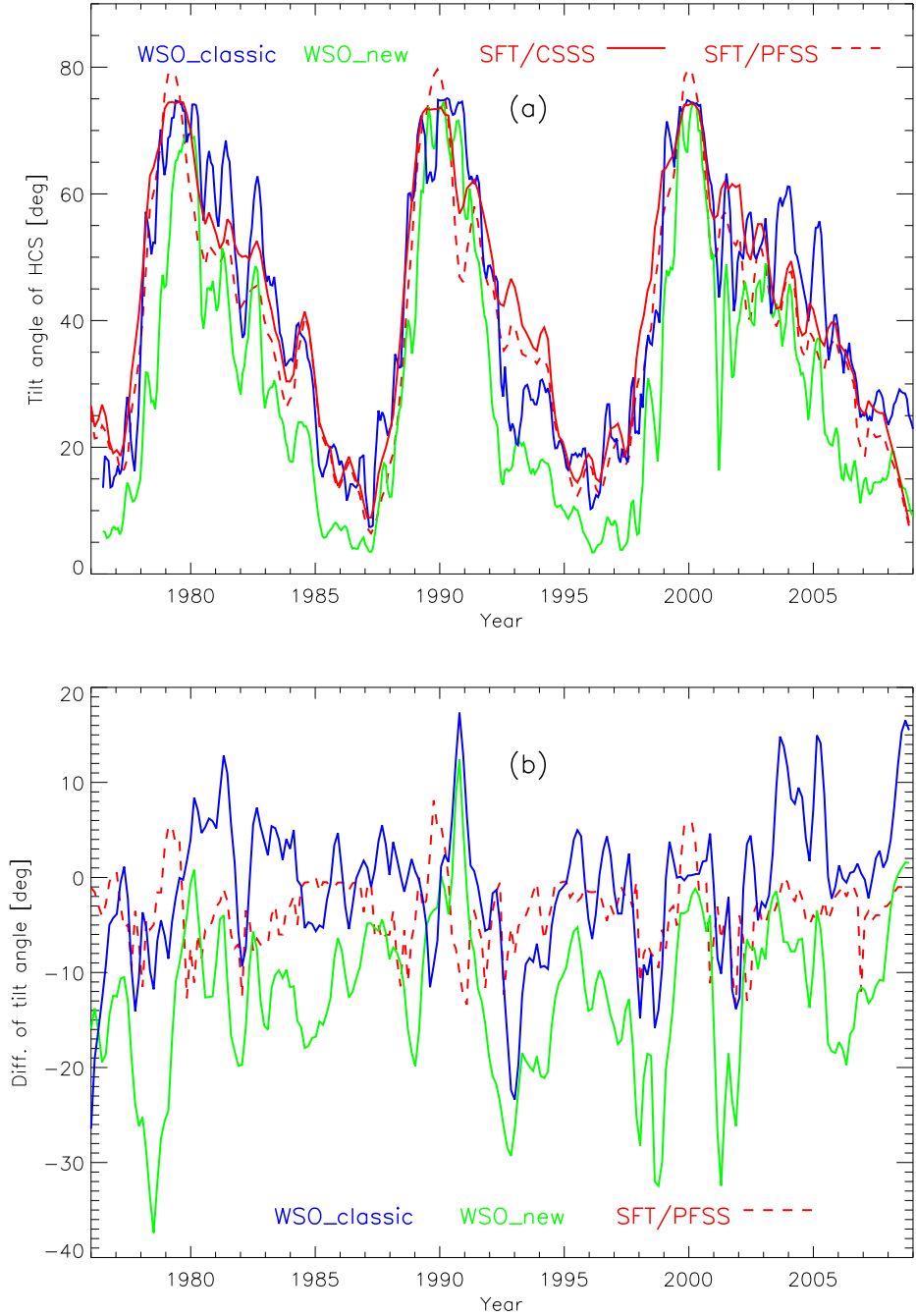} \caption{(a) Temporal evolution of
the HCS tilt angle. Solid red curve: SFT result with CSSS
extrapolation ($a=0.2R_\odot, R_{\rm{cs}=1.8R_\odot,
R_{\rm{ss}}=10R_\odot}$); dashed red curve: SFT result with PFSS
extrapolation ($R_{\rm{ss}}=3.25R_\odot$); blue curve: PFSS
extrapolation of WSO synoptic maps with line-of-sight field boundary
condition (`classic' model, $R_{\rm{ss}}=2.5R_\odot$); green curve:
PFSS extrapolation of WSO synoptic maps with radial field boundary
condition (`new' model, $R_{\rm{ss}}=3.25R_\odot$). The data for the
blue and green curves have been obtained from the WSO website
(http://wso.stanford.edu/Tilts.html). (b) The difference between the
HCS tilt angle from the SFT/CSSS and the three other methods shown
in the same color scheme as used in (a).}
\end{figure}

%% If you are not including electonic art with your submission, you may
%% mark up your captions using the \figcaption command. See the
%% User Guide for details.
%%
%% No more than seven \figcaption commands are allowed per page,
%% so if you have more than seven captions, insert a \clearpage
%% after every seventh one.

%% Tables should be submitted one per page, so put a \clearpage before
%% each one.

%% Two options are available to the author for producing tables:  the
%% deluxetable environment provided by the AASTeX package or the LaTeX
%% table environment.  Use of deluxetable is preferred.
%%

%% Three table samples follow, two marked up in the deluxetable environment,
%% one marked up as a LaTeX table.

%% In this first example, note that the \tabletypesize{}
%% command has been used to reduce the font size of the table.
%% We also use the \rotate command to rotate the table to
%% landscape orientation since it is very wide even at the
%% reduced font size.
%%

%% If you use the table environment, please indicate horizontal rules using
%% \tableline, not \hline.
%% Do not put multiple tabular environments within a single table.
%% The optional \label should appear inside the \caption command.

\clearpage

%% The following command ends your manuscript. LaTeX will ignore any text
%% that appears after it.


\begin{thebibliography}{}

\bibitem[Alanko-Huotari et al.(2007)]{alan07} Alanko-Huotari, K.,
Usoskin, I. G., Mursula, K., \& Kovaltsov, G. A. 2007, Adv. Space
Res., 40, 1064
\bibitem[Altschuler \& Newkirk(1969)]{alt69} Altschuler, M. D., \&
Newkirk, G. 1969, Sol. Phys., 9, 131
\bibitem[Arge et al.(2002)]{arg02} Arge, C. N., Hildner, E., Pizzo, V. J.,
\& Harvey, J. W. 2002, J. Geophys. Res., 107, 1319
\bibitem[Balogh et al.(1995)]{bal95} Balogh, A., Smith, E. J.,
 Tsurutani, B. T., Southwood, D. J., Forsyth, R. J., \& Horbury, T. S.
 1995, Science, 268, 1007
\bibitem[Banaszkiewicz et al.(1998)]{bana98} Banaszkiewicz, M.,
Axford, W. I., \& McKenzie, J. F. 1998, A\&A, 337, 940
\bibitem[Baumann et al.(2004)]{bau04} Baumann, I., Schmitt, D.,
Sch\"{u}ssler, M., \& Solanki, S. K. 2004, A\&A, 426, 1075
\bibitem[Baumann et al.(2006)]{bau06} Baumann, I., Schmitt, D., \&
Sch\"{u}ssler, M. 2006, A\&A, 446, 307
\bibitem[Beer(2000)] {beer00} Beer, J. 2000, Space Sci. Rev., 94, 53
\bibitem[Bogdan \& Low(1986)]{bog86} Bogdan, T. J., \& Low, B. C.
1986, \apj, 306, 271
\bibitem[Chapman et al.(1997)]{cha97} Chapman, G. A., Cookson, A. M., \&
Dobias, J. J. 1997, \apj, 482, 541
\bibitem[Ferreira \& Potgieter(2004)]{fer04} Ferreira, S. E. S., \&
Potgieter, M. S. 2004, \apj, 603, 744
\bibitem[Gizon \& Duvall(2004)]{giz04} Gizon, L. \& Duvall, T. L.
2004, in Multi-Wavelength Investigations of Solar Activity, ed.
Stepanov, A. V., Benevolenskaya, E. E., \& Kosovichev, A. G., IAU
Symp., 223, 41
\bibitem[Heber et al.(2009)]{heb09} Heber, B., Kopp, A., Gieseler, J.,
 M\"{u}ller-Mellin, R., Fichtner, H., Scherer, K., Potgieter, M. S., \&
  Ferreira, S. E. S. 2009, \apj, 699, 1956
\bibitem[Hoeksema et al.(1982)]{hoe82} Hoeksema, J. T.,
 Wilcox, J. M., \& Scherrer, P. H. 1982, J. Geophys. Res., 87, 10331
\bibitem[Howard(1991)]{howa91} Howard, R. F. 1991, Sol. Phys., 136, 251
\bibitem[Hu et al. (2008)]{hu08} Hu, Y. Q., Feng, X. S., Wu, S. T.,
\& Song, W. B. 2008, J. Geophys. Res., 113, A03106
\bibitem[Jiang et al.(2009)]{jia09} Jiang, J, Cameron, R., Schmitt, D., \&
Sch\"{u}ssler, M. 2009, \apj, 693, L96
\bibitem[K\'{o}ta \& Jokipii(1983)]{kot83} K\'{o}ta, J., \& Jokipii,
J. R. 1983, \apj, 265, 573
\bibitem[Leighton(1964)]{lei64} Leighton, R. B. 1964, \apj, 140, 1547
\bibitem[Lockwood et al.(2006)]{loc06} Lockwood, M., Rouillard, A. P.,
Finch, I., \& Stamper, R. 2006, J. Geophys. Res., 111, A09109
\bibitem[Lockwood et al.(2009)]{lock09} Lockwood, M., Rouillard, A. P., \&
 Finch, I. D. 2009, \apj, 700, 937
\bibitem[Mackay et al.(2002)]{mac02} Mackay, D. H.,
Priest, E. R., \& Lockwood, M. 2002, Sol. Phys., 209, 287
\bibitem[Marsch \& Richter (1984)]{mar84} Marsch, E., \& Richter, A.
K. 1984, J. Geophys. Res., 89, 5386
\bibitem[Owens et al.(2008)]{owens08} Owens, M. J., Arge, C. N.,
Crooker, N. U., Schwadron, N. A., \& Horbury, T. S. 2008, J.
Geophys. Res., 113, 12103
\bibitem[Petrie \& Patrikeeva (2009)]{pet09} Petrie, G. J. D., \&
Patrikeeva, I. 2009, \apj, 699, 871
\bibitem[Pulkkinen(2007)]{pulk07} Pulkkinen, T. 2007, Living Rev.
Solar Phys., 1, 4
\bibitem[Riley et al.(2002)]{ril02} Riley, P., Linker, J. A., \&
Miki\'{c}, Z. 2002, J. Geophys. Res., 107, 1136
\bibitem[Schatten(2005)]{schat05} Schatten, K., 2005,
Geophys. Res. Lett., 32, 21106
\bibitem[Schatten et al.(1969)]{sch69} Schatten, K. H.,
Wilcox, J. M., \& Ness, N. F. 1969, Sol. Phys., 6, 442
\bibitem[Sch\"{u}ssler \& Baumann(2006)]{sch06} Sch\"{u}ssler, M.,
\& Baumann, I. 2006, A\&A, 459, 945
\bibitem[Schrijver \& Liu(2008)]{sch08} Schrijver, C. J., \& Liu, Y. 2008,
Sol. Phys., 252, 19
\bibitem[Sivaraman et al.(1999)]{siva99} Sivaraman, K. R., Gupta, S. S.,
\& Howard, R. F. 1999, Sol. Phys., 189, 69
\bibitem[Snodgrass(1983)] {sno83} Snodgrass, H. B. 1983, \apj,
    501, 866
\bibitem[Solanki(1993)] {sol93} Solanki, S. K., 1993, Space Sci. Rev., 63, 1
\bibitem[Svalgaard et al.(2005)] {sval05}
    Svalgaard, L., Cliver, E. W. \& Kamide, Y. 2005,
    Geophys. Res. Lett., 32, L01104
\bibitem[van Ballegooijen et al.(1998)] {van98} van Ballegooijen, A. A.,
    Cartledge, N. P., \& Priest, E. R., 1998, \apj, 501, 866
\bibitem[Wang et al.(1989a)]{wang89a} Wang, Y.-M., Nash, A. G., \&
Sheeley, N. R., Jr. 1989a, Sol. Phys., 347, 529
\bibitem[Wang et al.(1989b)]{wang89b} Wang, Y.-M., Nash, A. G., \&
Sheeley, N. R., Jr. 1989b, Science, 245, 712
\bibitem[Wang \& Sheeley(1992)]{wan92} Wang, Y.-M., \& Sheeley, N. R.,
Jr. 1992, \apj, 392, 310
\bibitem[Wang et al.(2000)]{wang00} Wang, Y.-M, Sheeley,
N. R., Jr., \& Lean, J. 2000, Geophys. Res. Lett., 27, 621
\bibitem[Wang et al.(2002)]{wan02} Wang, Y.-M., Sheeley, N. R., Jr., \&
Lean, J. 2002, \apj, 580, 1188
\bibitem[Zhao \& Hoeksema(1995a)]{zha95a} Zhao, X. P., Hoeksema, J.
T. 1995a, J. Geophys. Res., 100, 19
\bibitem[Zhao \& Hoeksema(1995b)]{zha95b} Zhao, X. P., Hoeksema, J.
T. 1995b, Space Sci. Rev., 72, 189
\bibitem[Zhao et al.(2002)]{zha02} Zhao, X. P., Hoeksema, J.
T., \& Rich, N. R. 2002, Adv. Space Res., 29, 411
\end{thebibliography}
\end{document}